\documentstyle[12pt,epsf]{article}
\begin{document}

\begin{flushright}
SLAC-PUB-8037 \\ 
IITAP-98-010 \\
hep-ph/9901229 \\
December 1998 \\
\end{flushright}

\vspace*{0.5cm}
\begin{center}
{\Large \bf The QCD Pomeron with
Optimal Renormalization}\footnote{This work was supported in part by the Russian 
Foundation for Basic Research (RFBR): Grant Nos. 96-02-16717, 96-02-18897, 
98-02-17885;  INTAS: Grant No. 1867-93; INTAS-RFBR: Grant No. 95-0311; 
CRDF: Grant No. RP1-253; and the U.S. Department of Energy: Contract 
No.  DE-AC03-76SF00515.}

\end{center}
\vspace*{0.5cm}
\begin{center}
{\large Stanley~J.~Brodsky${}^{\ast}$, Victor~S.~Fadin${}^{\dagger}$,
Victor~T.~Kim${}^{\ddagger \&}$, Lev~N.~Lipatov${}^{\ddagger}$ \\
and \\
Grigorii~B.~Pivovarov${}^{\S \&}$ }
\end{center}
\vspace*{0.5cm} 
{${}^{\ast}$ : Stanford Linear Accelerator Center, Stanford University,
Stanford, CA 94309, USA \newline
${}^\dagger$ : Budker Institute of Nuclear Physics, 630090 Novosibirsk,
Russia \newline
${}^\ddagger$ : St.Petersburg Nuclear Physics Institute,  188350 Gatchina,
Russia \newline
${}^\S$ : Institute for Nuclear Research, 117312 Moscow, Russia \newline
${}^\&$ : International Institute of Theoretical and Applied Physics, Ames,
IA 50011, USA}
\vspace*{1cm}

\begin{center}
{\large \bf Abstract}
\end{center}

It is shown that the  next-to-leading order (NLO) corrections to the 
QCD Pomeron intercept obtained from the BFKL equation, when evaluated 
in non-Abelian physical renormalization schemes with BLM optimal scale 
setting do not exhibit the serious problems encountered in the 
$\overline{\mbox{MS}}$-scheme. 
A striking feature of the NLO BFKL Pomeron intercept in the BLM approach 
is its rather weak dependence on the virtuality of the reggeized gluon.
This remarkable property yields an important approximate conformal invariance. 
The results obtained provide an opportunity for applications of NLO BFKL 
resummation to high-energy phenomenology.

\vspace*{1cm}
PACS number(s): 12.38Cy, 12.40Nn
\vspace*{2cm}

\begin{center}
{\sl Brief version published in  Pis'ma ZhETF {\bf 70}, 161 (1999) 
[JETP Letters {\bf 70}, 155 (1999)]}
\end{center}

%PACS number(s): 12.38Cy, 12.40Nn

\newpage

The discovery of rapidly increasing structure functions in deep inelastic
scattering (DIS) at HERA \cite{HERA} at small-x is in agreement with the
expectations of the QCD high-energy limit.  
The Balitsky-Fadin-Kuraev-Lipatov (BFKL) \cite{BFKL} resummation of energy 
logarithms is anticipated to be an important tool for exploring this limit. 
The leading order (LO) BFKL calculations \cite{BFKL} predict a steep
rise of QCD cross sections. 
Namely, the highest eigenvalue, $\omega^{max}$, of the BFKL 
equation \cite{BFKL} is 
related to the intercept of the Pomeron which in turn governs 
the high-energy asymptotics of the cross sections: $\sigma \sim 
s^{\alpha_{I \negthinspace P}-1} = s^{\omega^{max}}$. 
The BFKL Pomeron intercept in the LO turns out to be rather large: 
$\alpha_{I \negthinspace P} - 1 =\omega_L^{max} = 
12 \, \ln2 \, ( \alpha_S/\pi )  \simeq 0.55 $ for 
$\alpha_S=0.2$; hence, it is very important to know the next-to-leading order
(NLO) corrections.
In addition, the LO BFKL calculations have restricted phenomenological 
applications because, {\it e.g.}, the running  of the QCD coupling 
constant $\alpha_S$ is not included, and the kinematic range of 
validity of LO BFKL is not known.

Recently the  NLO corrections to the BFKL resummation
of energy logarithms were calculated; see Refs. \cite{FL,CC98} and references
therein. The NLO corrections \cite{FL,CC98} to the highest eigenvalue 
of the BFKL equation turn out to be negative and even larger
than the LO contribution for $\alpha_S > 0.157$. In such circumstances
the phenomenological significance of the NLO BFKL calculations seems to be
rather obscure.

However, one should stress that the NLO calculations,
as any finite-order perturbative results, contain 
both  renormalization scheme and renormalization scale ambiguities.
The NLO BFKL calculations \cite{FL,CC98} were 
performed by employing the modified minimal subtraction scheme 
($\overline{\mbox{MS}}$) \cite{Bar78} to regulate the ultraviolet 
divergences with arbitrary scale setting.

In this work we consider the NLO BFKL resummation of energy logarithms 
\cite{FL,CC98} in physical renormalization schemes in order to study the
renormalization scheme dependence. To resolve the renormalization scale 
ambiguity we utilize  Brodsky-Lepage-Mackenzie (BLM) optimal scale 
setting \cite{BLM}. 
We show that the reliability of QCD predictions for the intercept of the
BFKL Pomeron at NLO when evaluated using  BLM scale setting 
within non-Abelian
physical schemes, such as the momentum space 
subtraction (MOM) scheme \cite{Cel79,Pas80} 
or the $\Upsilon$-scheme based on  $\Upsilon \rightarrow ggg$ decay, 
is significantly improved compared to the $\overline{\mbox{MS}}$-scheme. 
This provides a basis for applications of NLO BFKL resummation to 
high-energy phenomenology.

We begin with the representation of the $\overline{\mbox{MS}}$-result
of NLO BFKL \cite{FL,CC98} in physical renormalization schemes.
Although the $\overline{\mbox{MS}}$-scheme is somewhat artificial and 
lacks a clear physical picture, it 
can serve as a convenient intermediate renormalization 
scheme. The eigenvalue of the NLO BFKL equation  at transferred momentum 
squared $t=0$ in the $\overline{\mbox{MS}}$-scheme 
\cite{FL,CC98} can be represented as the action of 
the NLO BFKL kernel (averaged over azimuthal angle) on the LO 
eigenfunctions $(Q_{2}^{2}/Q_{1}^{2})^{-1/2+i\nu }$  \cite{FL}: 
\begin{eqnarray}
\omega _{\overline{MS}}(Q_{1}^{2},\nu ) &=&\int d^{2}Q_{2}\,\,\,
K_{\overline{MS}}(\vec{Q_{1}},\vec{Q_{2}})
\left( \frac{Q_{2}^{2}}{Q_{1}^{2}}
\right) ^{-\frac{1}{2}+i\nu }=  \nonumber \\
&=&  N_C \chi_{L}(\nu ) \frac{\alpha_{\overline{MS}}(Q_{1}^{2})}{\pi }
\Biggl[ 1 + r_{\overline{MS}}(\nu ) 
\frac{\alpha_{\overline{MS}}(Q_{1}^{2})}{\pi } \Biggr] ,
\label{kernelact} 
\end{eqnarray}
where 
\[
\chi _{L}(\nu )=2\psi (1)-\psi (1/2+i\nu)-\psi (1/2-i\nu)
\]
is the function related with the LO eigenvalue, 
$\psi =\Gamma ^{\prime}/\Gamma $ denotes the Euler $\psi $-function, the 
$\nu $-variable is conformal weight parameter \cite{Lipatov97}, $N_C$ is
the number of colors, and $Q_{1,2}$ are the virtualities 
of the reggeized gluons.

The calculations of Refs. \cite{FL,CC98} allow us
to decompose the NLO coefficient $r_{\overline{MS}}$ of 
Eq. (\ref{kernelact}) into $\beta$-dependent and 
the conformal ($\beta$-independent) parts:   
\begin{equation}
r_{ \overline{MS}} (\nu) = 
r_{ \overline{MS}}^{\beta}(\nu) + r_{ \overline{MS}}^{conf} (\nu) ,
\label{evnl}
\end{equation}
where 
\begin{equation}
r_{ \overline{MS}}^{\beta}(\nu) 
 = - \frac{\beta_0}{4} \Biggl[ 
\frac{1}{2} \chi_{L}(\nu ) - \frac{5}{3} \Biggr]
\label{chimsbeta}
\end{equation}
and  
\begin{eqnarray}
r_{\overline{MS}}^{conf} (\nu) 
& = &  - \frac{N_C}{4 \chi_{L}(\nu )} \left[ 
\frac{\pi^2 \sinh (\pi \nu )} 
{2 \nu \cosh^2 (\pi \nu )} 
\left( 3 + \left( 1+ \frac{N_F}{N_C^3} \right) \right. 
\frac{11 + 12 \nu^2 }
{16 (1 + \nu^2 )} \right) - \chi_{L}^{\prime \prime }(\nu ) 
 \nonumber \\
& & \left. +  \frac{\pi^2-4}{3} \chi_{L}(\nu ) - 
\frac{\pi^3}{\cosh(\pi \nu )} - 6\zeta(3) + 4\varphi(\nu) \right] \, 
\label{chimsconf}
\end{eqnarray}
with 
\begin{equation}
\varphi (\nu ) = 2 \int_0^1 dx \frac{\cos(\nu \ln(x))}{(1+x) \sqrt{x}} 
\Biggl[ \frac{\pi^2}{6} - {\mathrm Li}_2 (x) \Biggr], \; 
{\mathrm Li}_2 (x) = - \int_0^x dt \frac{\ln (1-t)}{t}.
\end{equation}
Here $\beta_0 = (11/3)N_C - (2/3) N_{F}$ is the leading coefficient
of the QCD $\beta$-function, $N_F$ is the number of flavors,
$\zeta(n)$ stands for the Riemann zeta-function, 
${\mathrm Li}_2 (x)$ 
is the Euler dilogarithm (Spence-function).
In Eq. (\ref{chimsconf}) $N_F$  denotes flavor number of 
the Abelian part of the 
$gg \rightarrow q\overline{q}$ process contribution. 
The Abelian part is not associated with the running of the
coupling \cite{BH} and is consistent with the correspondent QED 
result for the $\gamma^{\ast} \gamma^{\ast} \rightarrow e^+e^- $ 
cross section \cite{GLF}.

The $\beta$-dependent NLO coefficient  $r_{ \overline{MS}}^{\beta} (\nu)$, 
which is related to the running of the coupling, 
receives contributions from the gluon reggeization diagrams, 
from the virtual part of the one-gluon emission, from the real 
two-gluon emission, and from the non-Abelian part \cite{BH} of the 
$gg \rightarrow q\overline{q}$ process.
There is an omitted term in $r_{\overline{MS}}^{\beta} (\nu )$ 
proportional to $\chi^{\prime}_L(\nu)$ which originates
from the asymmetric treatment of $Q_1$ and $Q_2$
and which can be removed by the redefinition of the LO
eigenfunctions \cite{FL}.

The NLO BFKL Pomeron intercept then  reads for $N_C=3$: \cite{FL} 
\begin{equation}
\alpha_{I \negthinspace P}^{\overline{MS}} - 1  = 
\omega_{\overline{MS}}(Q^2,0) =
12 \, \ln2 \, \frac{ \alpha_{\overline{MS}}(Q^2)}{\pi} \biggl[ 
1 + r_{\overline{MS}}(0) 
\frac{\alpha_{\overline{MS}}(Q^2)}{\pi} \biggr] \, , 
\end{equation} 
\begin{equation}
r_{\overline{MS}}(0) \simeq -20.12 - 0.1020 N_F + 0.06692 \beta_0 ,
\label{rms0}
\end{equation}
$$r_{\overline{MS}}(0)_{\vert N_F =4} \simeq -19.99 .$$

Physical renormalization schemes provide small and physically meaningful 
perturbative coefficients by incorporating large corrections into the 
definition of the coupling constant. 
One of the most popular physical schemes
is MOM-scheme 
\cite{Cel79,Pas80}, based on renormalization of the triple-gluon vertex 
at some symmetric off-shell momentum. However, in the MOM-scheme
the coupling constant is gauge-dependent already in the LO, and  there
are rather cumbersome technical difficulties. These 
difficulties can be avoided by performing calculations in the intermediate 
$\overline{\mbox{MS}}$-scheme, and then by making the transition to some
physical scheme by a finite renormalization \cite{Cel79}.
In order to eliminate the dependence on gauge choice and other theoretical 
conventions, one can consider renormalization 
schemes based on physical processes \cite{BLM}, {\it e.g.}, 
V-scheme based on heavy quark potential.
Alternatively, one can introduce a physical scheme based on 
$\Upsilon \rightarrow ggg$ decay using the NLO calculations of Ref. \cite{ML}.

A finite renormalization due to the change of scheme can be accomplished by
a transformation of the QCD coupling \cite{Cel79}:
\begin{equation}
\alpha_{S} \rightarrow \alpha_{S} \biggl[ 1 + T \frac{\alpha_{S}}{\pi} 
\biggr] ,
\end{equation}
where T is some function of $N_C$, $N_F$, and for the MOM-scheme, of
a gauge parameter $\xi$. 
Then the NLO BFKL eigenvalue in 
the MOM-scheme can be represented  as follows 
\begin{eqnarray}
\omega_{MOM}(Q^2,\nu) 
& = & N_C \chi_{L}(\nu) \frac{\alpha_{MOM} (Q^2)}{\pi} \biggl[ 1 +
r_{MOM}(\nu)\,\, \frac{\alpha_{MOM}(Q^2)} {\pi}\biggr] \, , \\
r_{MOM}(\nu) & = & r_{\overline{MS}} (\nu) + T_{MOM} . \nonumber
\end{eqnarray}
For the transition from the $\overline{\mbox{MS}}$-scheme to the
MOM-scheme the corresponding T-function has the following form 
\cite{Cel79}: 
\begin{eqnarray}
T_{MOM}& = & T_{MOM}^{conf}+T_{MOM}^{\beta}, \\
T_{MOM}^{conf} &=& \frac{N_C}{8} \biggl[ \frac{17}{2} I + 
\xi \frac{3}{2} (I-1) + \xi^2 (1-\frac{1}{3}I) - 
\xi^3 \frac{1}{6} \biggr]  , \nonumber \\
T_{MOM}^{\beta} &=& - \frac{ \beta_0}{2} \biggl[ 1 +\frac{2}{3} I \bigg] , 
\nonumber
\end{eqnarray}
where $I=-2 \int^{1}_{0}dx \ln(x)/[x^2-x+1]\simeq 2.3439$.

Analogously, one can obtain for the V-scheme \cite{BLM}: 
\begin{equation}
T_V  =   \frac{2}{3} N_C -\frac{5}{12} \beta_0, 
\end{equation}
and by the use of the results of Ref. \cite{ML} for the $\Upsilon$-scheme:
\begin{equation}
T_{\Upsilon}  = \frac{6.47}{3} N_C - \frac{2.77}{3} \beta_0. 
\end{equation}

\begin{figure}[htb]
\begin{center}
\leavevmode
{\epsfxsize=8.5cm\epsfysize=8.5cm\epsfbox{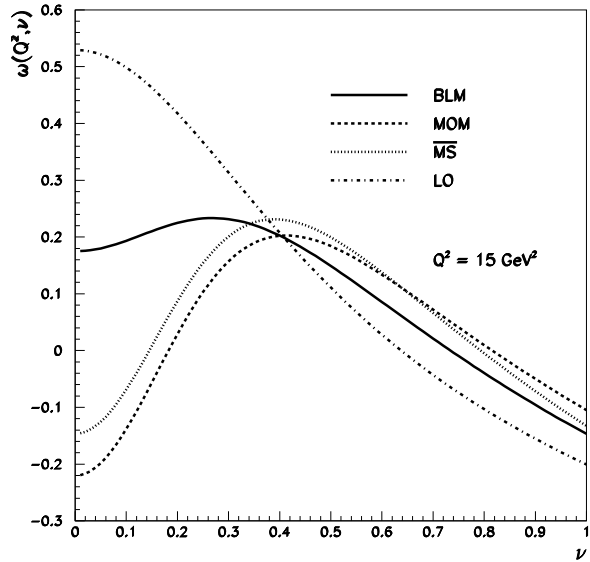}}
\end{center}
\caption[*]{$\nu$-dependence of the NLO BFKL eigenvalue at $Q^2=15$
GeV$^2$: BLM (in MOM-scheme) -- solid, 
MOM-scheme (Yennie gauge: $\xi=3$) -- dashed, 
$\overline{\mbox{MS}}$-scheme -- dotted. LO
BFKL ($\alpha_S=0.2$) -- dash-dotted.}
\label{fig:1}
\end{figure}

One can see from Table \ref{fig:1} that the problem of a large NLO
BFKL coefficient remains. The large size of the perturbative
corrections leads to significant renormalization scale ambiguity.

% Table 1
\begin{table}
\begin{center}
\begin{tabular}{|c|c|c|c|c|}
\hline
\multicolumn{2}{|c|}{Scheme}  & $T=T^{conf}+T^{\beta}$ &
 $r(0)=r^{conf}(0)+r^{\beta}(0)$
& $r(0)$        \\ 
\multicolumn{2}{|c|}{} & & & $(N_F = 4)$ \\ 
\hline\hline
M & $\xi=0$ & $7.471 - 1.281 \beta_0$ & 
$-12.64 - 0.1020 N_F - 1.214 \beta_0$ & -22.76 \\
 \cline{2-5}
O & $\xi=1$ & $8.247 - 1.281 \beta_0$ & 
$-11.87 - 0.1020 N_F - 1.214 \beta_0$ & -21.99 \\  
\cline{2-5}
M & $\xi=3$ & $8.790 -1.281 \beta_0$ & 
$-11.33 - 0.1020 N_F - 1.214 \beta_0$& -21.44  \\ 
\hline
\multicolumn{2}{|c|}{V} & $ 2 - 0.4167 \beta_0 $ & 
$-18.12 - 0.1020 N_F - 0.3497 \beta_0$ & -21.44 \\ \hline
\multicolumn{2}{|c|}{$\Upsilon$} & $6.47 - 0.923 \beta_0$ & 
$-13.6 - 0.102 N_F - 0.856 \beta_0 $ & -21.7  \\ 
\hline
\end{tabular}
\end{center}
\caption{Scheme-transition function and the NLO BFKL coefficient 
in physical schemes.}
\label{tab:1}
\end{table}

The renormalization scale ambiguity problem can be resolved if one can
optimize the choice of scales and renormalization schemes according to some
sensible criteria. In the BLM optimal scale
setting \cite{BLM}, the renormalization scales are chosen such that
all vacuum polarization effects from the QCD $\beta$-function are resummed
into the running couplings. The coefficients of the perturbative series are
thus identical to the perturbative coefficients of the corresponding
conformally invariant theory with $\beta=0$. The BLM  
approach has the important advantage of resumming the large and strongly
divergent terms in the perturbative QCD series which grow as $n! [\alpha_S
\beta_0 ]^n$, {\it i.e.}, the infrared renormalons
associated with coupling constant renormalization.
The renormalization scales in the BLM approach are physical in
the sense that they reflect the mean virtuality of the gluon 
propagators \cite{BLM}. 

BLM scale setting \cite{BLM} can be applied within any
appropriate physical scheme.
In the present case one can show that within the 
V-scheme (or the $\overline{\mbox{MS}}$-scheme) 
the BLM procedure does not change significantly the value of 
the NLO coefficient $r(\nu)$. 
This can be understood since the V-scheme, as well as 
$\overline{\mbox{MS}}$-scheme, are adjusted primarily to
the case when in the LO there are dominant QED (Abelian) type
contributions, whereas in the BFKL case there are important LO 
gluon-gluon (non-Abelian) interactions. 

Therefore, from the point of view of BLM scale setting, one
can separate QCD processes into two classes specifying whether gluons are
involved to the LO or not.
Thus one can expect that in the BFKL case 
it is appropriate to use a physical scheme which is adjusted for
non-Abelian interactions in the LO. One can choose the MOM-scheme
based on  the symmetric triple-gluon 
vertex \cite{Cel79,Pas80} or the $\Upsilon$-scheme based on 
$\Upsilon \rightarrow ggg$ decay.
The importance of taking into account this circumstance
for vacuum polarization effects one can be seen from the 
``incorrect'' sign of the $\beta_0$-term for $r_{\overline{MS}}$  
in the unphysical $\overline{\mbox{MS}}$-scheme (Eq. (\ref{rms0})).

Adopting  BLM scale setting, the NLO BFKL eigenvalue
in the MOM-scheme is 
\begin{equation}
\omega_{BLM}^{MOM}(Q^{2},\nu) =  
N_C \chi_{L} (\nu) \frac{\alpha_{MOM}(Q^{MOM \, 2}_{BLM})}{\pi}
\Biggl[1 + 
r_{BLM}^{MOM} (\nu) \frac{\alpha_{MOM}(Q^{MOM \, 2}_{BLM})}{\pi} \Biggr] ,
\end{equation}
\begin{equation}
r_{BLM}^{MOM} (\nu) = r_{MOM}^{conf} (\nu) \, .
\end{equation}

The $\beta$-dependent part of the $r_{MOM}(\nu)$ defines the
corresponding BLM optimal scale 
\begin{equation}
Q^{MOM \, 2}_{BLM} (\nu) = Q^2 \exp 
\Biggl[ - \frac{4 r_{MOM}^{\beta}(\nu)}{\beta_0} \Biggr] 
= Q^2 \exp \Biggl[ \frac
{1}{2}\chi_L (\nu) - \frac{5}{3} + 2 \biggl(1+\frac{2}{3} I \biggr) \Biggr].
\label{qblm}
\end{equation}
Taking into account the fact that $\chi_L(\nu) \rightarrow -2 \ln (\nu) $ at 
$\nu \rightarrow \infty $, one obtains at large $\nu$ 
\begin{equation}
Q^{MOM \, 2}_{BLM} (\nu) = Q^2 \frac{1}{\nu} \exp \biggl[ 2 \biggl(1 + \frac{2}{3}
I \biggr) - \frac{5}{3} \biggr].
\end{equation}

At $\nu=0$ we
have $Q^{MOM \, 2}_{BLM} (0) = Q^2 \bigl( 4 \exp [2(1+2 I /3)-5/3] \bigr) \simeq
Q^2  \, 127$. Note that $Q^{MOM \, 2}_{BLM}(\nu)$ contains a large factor, 
$\exp [- 4 T_{MOM}^{\beta}/\beta_0 ] = \exp [2(1+2 I /3)] \simeq 168$, which 
reflects a large kinematic difference between MOM- and 
$\overline{\mbox{MS}}$- schemes \cite{Cel83,BLM}, even in an Abelian 
theory. 

% Table 2
\begin{table}
\begin{center}
\begin{tabular}{|c|c|c|c|c|c|}
\hline
\multicolumn{2}{|c|}{Scheme} & \multicolumn{1}{|c|}{$r_{BLM}(0)$} & 
\multicolumn{3}{|c|}
{$\alpha_{I \negthinspace P}^{BLM} - 1 =\omega_{BLM}(Q^2,0)$} \\ 
\cline{4-6}
\multicolumn{2}{|c|}{}  & \multicolumn{1}{|c|}{$(N_F = 4)$} & 
$Q^2=1$ GeV$^2$ & $Q^2=15$ GeV$^2$ & $Q^2=100$ GeV$^2$    \\ 
\hline\hline
M & $\xi=0$ & -13.05 & 0.134 & 0.155 & 0.157 \\
 \cline{2-6}
O & $\xi=1$ & -12.28 & 0.152 &  0.167 & 0.166 \\  
\cline{2-6}
M & $\xi=3$ & -11.74 & 0.165 & 0.175 & 0.173 \\ 
\hline
\multicolumn{2}{|c|}{$\Upsilon$} & -14.01 & 0.133  & 0.146 & 0.146 \\ 
\hline
\end{tabular}
\end{center}
\caption{The NLO BFKL Pomeron intercept in the BLM scale setting
within non-Abelian physical schemes.}
\label{tab:2}
\end{table}

Analogously one can implement the BLM scale setting in the
$\Upsilon$-scheme (Table \ref{tab:2}). 

% Fig. 1

\begin{figure}[htb]
\begin{center}
\leavevmode
{\epsfxsize=8.5cm\epsfysize=8.5cm\epsfbox{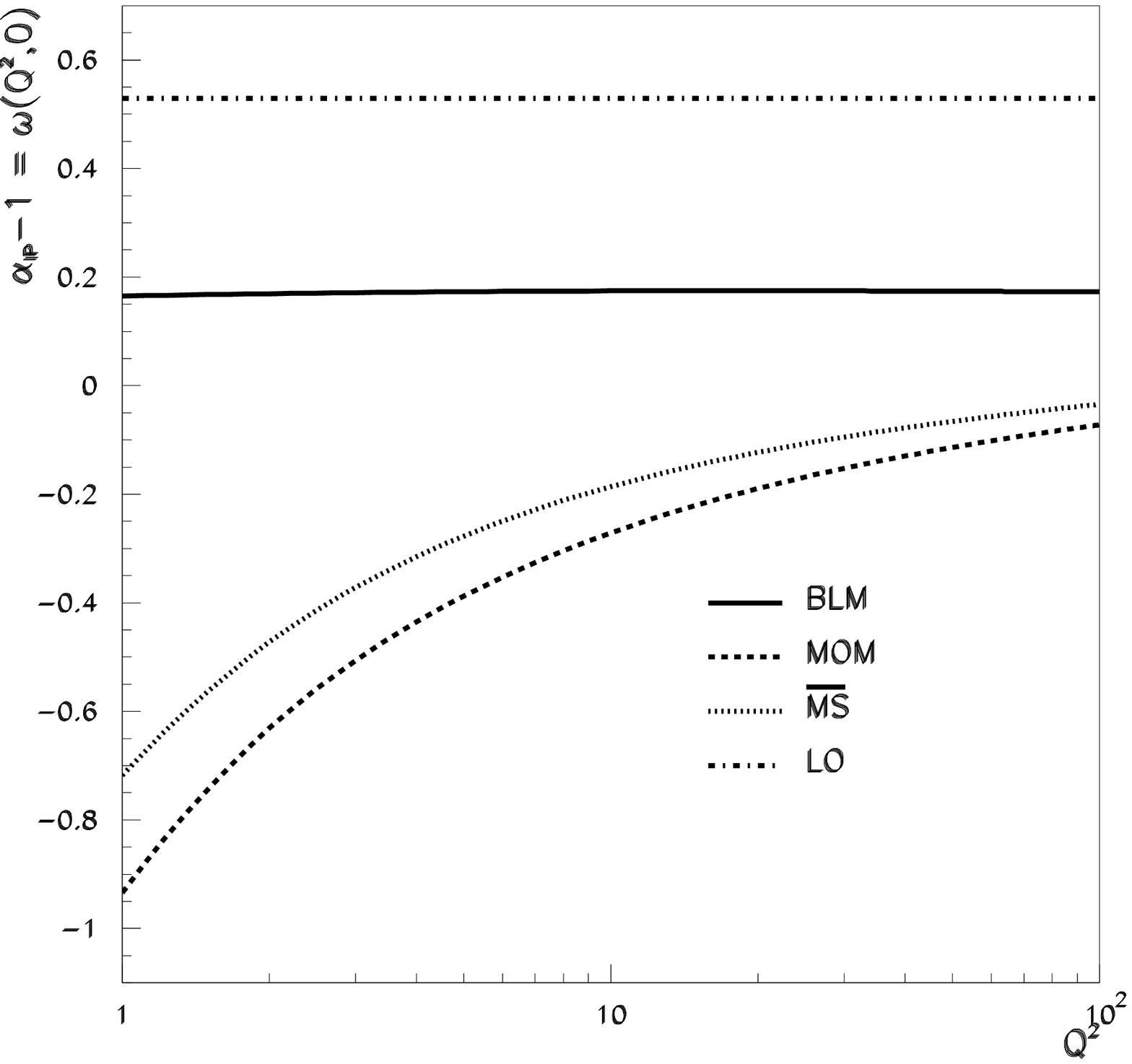}}
\end{center}
\caption[*]{$Q^2$-dependence of the BFKL Pomeron intercept in the NLO.
The notation is as in Fig. 1.}
\label{fig:2}
\end{figure}

Figs. \ref{fig:1} and \ref{fig:2} give the results for 
the eigenvalue of the NLO BFKL kernel. 
We have used the QCD parameter $\Lambda = 0.1$ GeV which
corresponds to $\alpha_S = 4 \pi / \bigl[ \beta_0 \ln(Q^2/\Lambda^2) \bigr] 
\simeq 0.2$ at $Q^2=15$ GeV$^2$. 
Also, the generalization \cite{BGMR,Shirkov} of the $\beta$
-function  in the running coupling and of flavor number
for continuous treatment of quark thresholds has been used.

% Fig. 2 

One can see from Fig. \ref{fig:1}, that the maximum which 
occurs at non-zero $\nu$ is
not as pronounced in the BLM approach compared to the 
$\overline{\mbox{MS}}$-scheme, and thus it will not serve 
as a good saddle point at high energies.

One of the striking features of this analysis is that the NLO value for 
the intercept of the BFKL Pomeron, improved by the BLM procedure, has a 
very weak dependence on the gluon virtuality $Q^2$. 
This agrees with the conventional Regge-theory where
one expects an universal intercept of the Pomeron without any $Q^2$-dependence.
The minor $Q^2$-dependence obtained, on one side, provides near insensitivity 
of the results to the precise value of $\Lambda$, and, on the other side, leads
to approximate scale and conformal invariance. Thus one may use conformal
symmetry \cite{Lipatov97,Lipatov86} for the continuation of the present results
to the case $t \neq 0$.

Therefore, by the applying of the BLM scale setting 
 within the non-Abelian physical schemes 
(MOM- and $\Upsilon$- schemes) 
we do not face the serious problems \cite{Ross,Kov98,Blu98} which were 
present in the 
$\overline{\mbox{MS}}$-scheme, {e.g.}, oscillatory cross 
section disbehavior based on the saddle point approximation 
\cite{Ross}, and the somewhat puzzling analytic structure 
\cite{Kov98} of the $\overline{\mbox{MS}}$-scheme result \cite{FL,CC98}.

Now we will briefly consider NLO BFKL within other approaches to the
optimization of perturbative theory, namely, fast apparent convergence (FAC) 
\cite{FAC} and the principle of minimal sensitivity (PMS) \cite{PMS}.

By the use of the FAC \cite{FAC} one can obtain 
\begin{equation}
\omega_{FAC}(Q^2,\nu) = N_C \chi_{L} (\nu)  
\frac{\alpha_{S}(Q^{2}_{FAC}(\nu))}{\pi},
\end{equation}
\begin{equation}
Q^{2}_{FAC} (\nu) = Q^2 \exp \Biggl[ - \frac{4}{\beta_0} r(\nu) \Biggr].
\end{equation}

In the $\overline{\mbox{MS}}$-scheme at $\nu=0$, 
$\omega_{FAC}=0.33-0.26$ for $Q^2=1-100$ GeV$^2$. However, the NLO coefficient $r(\nu)
$, and hence, the FAC effective scale, each have a singularity at 
$\nu_0 \simeq 0.6375$ due to a zero of the $\chi_L(\nu)$-function.

In the PMS approach \cite{PMS} the NLO BFKL eigenvalue reads as follows 

\begin{equation}
\omega_{PMS}(Q^2,\nu) = N_C\,\chi_{L} (\nu) \frac{\alpha_{PMS}(Q^{2}(\nu))}{\pi}
 \Biggl[ \frac{1+ (C/2) \alpha_{PMS}/\pi } {1+ C
\alpha_{PMS}/\pi} \Biggr],
\end{equation}
where the PMS effective coupling $\alpha_{PMS}$ is a solution of the
following transcendental equation 
\begin{equation}
\frac{\pi}{\alpha_{PMS}}+ C \ln \Biggl( \frac{C \alpha_{PMS}/\pi}
{ 1 + C \alpha_{PMS}/\pi} \Biggr) + 
\frac{C/2}{1 + C \alpha_{PMS}/\pi} = 
\frac{\beta_0}{4} \ln \Biggl( \frac{Q^2}{\Lambda^2}\Biggr) -
r(\nu)
\end{equation}
with $C=\beta_1/(4 \beta_0)$ and $\beta_1=102-38 N_F/3$. At $\nu=0$ 
one obtains in the $\overline{\mbox{MS}}$-scheme
$\omega_{PMS}=0.23-0.20$ for $Q^2=1-100$ GeV$^2$, 
but, by the same reason as in
the FAC case, the PMS effective coupling has a singularity at $\nu_0$.
Thus, the application of the FAC and PMS scale setting approaches to the
BFKL eigenvalue problem lead to difficulties with the conformal weight
dependence, an essential ingredient of BFKL calculations. 
The unphysical behavior of the FAC and PMS effective scales 
for jet production processes has been noted in Refs. \cite{Kra91}.

Before making conclusions a few remarks are in order. 

({\it i}) Since 
 the BFKL equation can be interpreted as the ``quantization'' 
 of a renormalization group
equation \cite{Lipatov86}, it follows that the effective scale 
should depend on the BFKL eigenvalue $\omega$, associated with 
the Lorentz spin, rather than on $\nu$. 
Thus, strictly speaking, one can use the above  
effective scales as function
of $\nu$ only in ``quasi-classical'' approximation at large-$Q^2$.
 However, the present remarkable $Q^2$-stabilty  leads us to expect that 
the results obtained with LO eigenfunctions may not change considerably
for $t \neq 0$ due to  the approximate conformal invariance.
This issue will be discussed in more detail in the extended version of this
work \cite{BFKLP}. 

({\it ii}) 
There have been a number of recent papers which analyze the NLO BFKL
predictions in terms of  rapidity correlations 
\cite{Lipatov98}, $t$-channel unitarity \cite{Coriano95},
angle-ordering \cite{CCFM},  double transverse momentum 
logarithms \cite{Salam} and  BLM scale setting 
for deep inelastic structure functions  \cite{Thorne}.
A discussion of these topics within our approach will be  
deferred to Ref. \cite{BFKLP}.

To conclude, we have shown that the NLO corrections to the BFKL 
equation for the QCD Pomeron become controllable
and meaningful provided one uses physical renormalization scales and
schemes relevant to non-Abelian gauge theory.  BLM optimal scale setting
automatically sets the appropriate physical renormalization scale by
absorbing the non-conformal $\beta$-dependent coefficients.   The
strong renormalization scheme dependence of the NLO corrections to BFKL
resummation then largely disappears.  This is in contrast to the unstable
NLO results obtained in the conventional $\overline{\mbox{MS}}$-scheme with
arbitrary choice of renormalization scale.
A striking feature of the NLO BFKL Pomeron intercept in 
the BLM approach is its very weak $Q^2$-dependence, which provides
approximate conformal invariance. 
The new results presented here open new windows for applications 
of NLO BFKL resummation to high-energy phenomenology.

\section*{Acknowledgement}

VTK, LNL and GBP are thankful to A.~R.~White for warm hospitality at the 
Argonne National Laboratory. VTK and GBP thank Fermilab Theory Group 
for their kind hospitality.

\end{document}